\documentclass{ws-mpla}

\begin{document}

\markboth{P.J. Boyle}
{The Elemental Composition of High-Energy Cosmic Rays: Measurements with TRACER}

\catchline{}{}{}{}{}

\title{The Elemental Composition of High-Energy Cosmic Rays: Measurements with TRACER}

\author{\footnotesize P.J. Boyle}
\address{Enrico Fermi Institute, The University of Chicago, 933 E. 56th Street \\
Chicago, Illinois 60637, USA. \\
boyle@uchicago.edu}


\maketitle

\pub{Received (Day Month Year)}{Revised (Day Month Year)}

\begin{abstract}

TRACER (``Transition Radiation Array for Cosmic Energetic
Radiation'') is a balloon borne instrument that has been developed to
directly measure the composition and energy spectra of individual
heavy elements up to 10$^{15}$~eV~particle$^{-1}$. TRACER achieves a
large geometric factor (5~m$^2$~sr) through the use of a Transition
Radiation Detector utilizing arrays of single wire proportional
tubes. TRACER has measured the energy spectra of the elements O, Ne,
Mg, Si, S, Ar, Ca, and Fe. The energy spectra reach energies in excess
of 10$^{14}$~eV~particle$^{-1}$ and exhibit nearly the same spectral
index (2.65 $\pm$ 0.05) for all elements.

\end{abstract}

\ccode{PACS Nos.: 95.55.Ym; 95.85.Ry; 98.70.Sa}

\section{Introduction}	
\begin{figure}[ht]
\begin{center}
\includegraphics[width=0.75\textwidth]{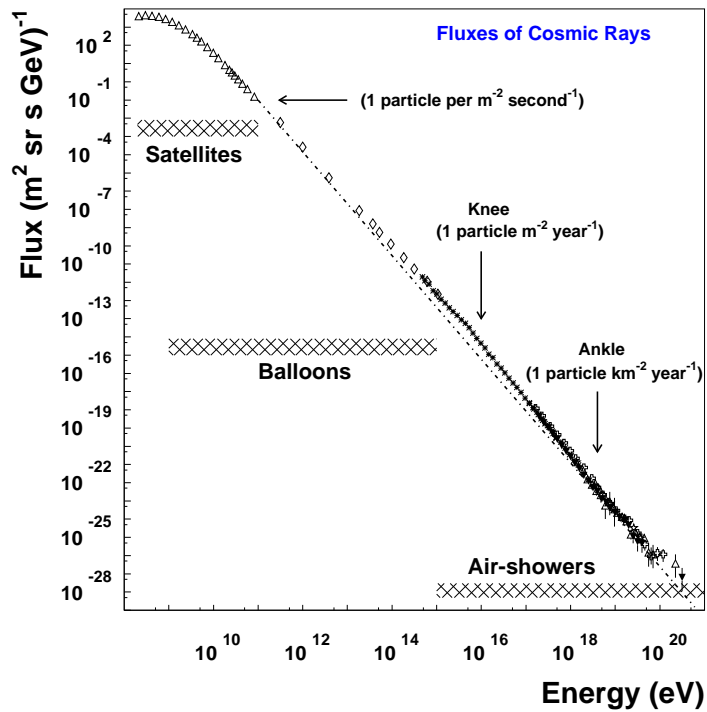}
\end{center}
\caption{All particle cosmic-ray spectrum. The hashed regions
indicates the energy regions sensitive to satellites, balloon based
detectors, and ground based air-shower detectors. [adapted from Cronin
et al.]}
\label{allparticle}
\end{figure}

Cosmic rays arriving at Earth span an energy range from 10$^{8}$~eV to
10$^{20}$~eV. Measurements of the cosmic-ray flux from experiments,
both on the ground and above the atmosphere, are presented in
Figure~\ref{allparticle} and show the cosmic-ray flux falling as a
near featureless power law over 30 decades. Accurate measurements of
the elemental composition and individual energy spectra has, however,
been an experimental challenge for many years. To date, few
measurements of the energy spectra above 10$^{12}$~eV~particle$^{-1}$
have been made for individual elements. In this paper we review an
effort by the TRACER group to directly measure the elemental
composition of cosmic rays up to 10$^{15}$~eV~particle$^{-1}$.

The TRACER (``Transition Radiation Array for Cosmic Energetic
Radiation'') instrument is a balloon borne detector and probably the
largest cosmic-ray detector ever flown above the atmosphere. TRACER
has had three successful balloon flights since 1999, yielding an
exposure of $\sim$70~m$^{2}$~sr~days. The observational goal of the
first two flights of TRACER was to measure the individual energy
spectra of the heavy elements O, Ne, Mg, Si, S, Ar, Ca, and Fe. For
the third flight the instrument was upgraded to be sensitive to the
light-medium elements B, C, and, N. Future flights of TRACER are
proposed and would include measurements of the the sub-Fe elements Sc,
Ti, V, Cr, and Mn.

This review begins with an overview of the TRACER concept and a
description of the instrument in Section~\ref{sec:tracer}. The
analysis of the TRACER data is detailed in Section~\ref{sec:analysis}
and the resulting energy spectra are presented and compared with
measurements from other experiments in Section~\ref{sec:results}. The
measurement of the elements B,C, and sub-Fe is discussed in
Section~\ref{sec:future}.

\section{The TRACER Concept}
\label{sec:tracer}

\begin{figure}[ht]
\begin{center}
\includegraphics[width=0.95\textwidth]{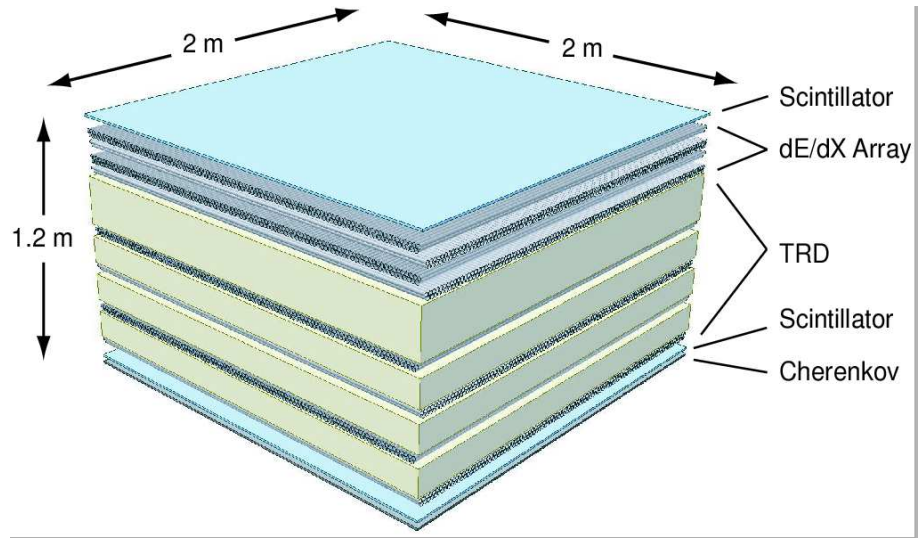}
\end{center}
\caption{Schematic diagram of the TRACER detector as flown during the
Antarctic balloon campaign in 2003. [From Ave et al.] }
\label{tracer}
\end{figure}

A successful cosmic-ray experiment must determine for each nucleus both
the nuclear charge \emph{Z} and the energy \emph{E} or Lorentz factor
$\gamma \approx E/mc^{2}$. TRACER realizes this goal by combining
scintillation and Cherenkov counters to measure the charge,
together with a \emph{dE/dx} counter and a Transition Radiation
Detector (TRD) to measure the energy of individual particles up to
10$^{15}$~eV~per~particle (see Figure~\ref{tracer}).

The key feature on TRACER is the employment of a TRD to measure the
energy or Lorentz factor of cosmic-ray nuclei up to
10$^{15}$~eV~per~particle. This application of TRD's is more
challenging than their use as threshold devices in accelerator or
cosmic-ray measurements [\refcite{tree},\refcite{ts93},\refcite{heat}],
where one just wishes to discriminate between particles of the same
energy but different mass (e.g. electrons and pions or protons).

Key features which distinguish TRD's from conventional energy
measuring devices such as magnet spectrometers or hadronic
calorimeters include: (a) a favorable area-to-weight ratio, which is
important for weight-restricted-instruments on balloons or spacecraft
(b) the possibility of accelerator calibrations with beams of
electrons or pions at large Lorentz factors, for which beams of nuclei
are not available (c) the fact that the TR-signal scales strictly with
$Z^{2}$. Therefore, relative signal fluctuations decrease as $1/Z$,
making possible increasingly precise measurements for particles with
higher charge. On the other hand, large signal fluctuations at low $Z$
restrict this use of TRD's to cosmic-ray nuclei heavier than helium
($Z \ge$ 3) (d) a good match between the Lorentz-factor range of a TRD
(400 - 10$^{5}$) and the range of energies that should be covered in
direct measurements (e) the possibility of redundant measurements in a
layered radiator/detector configuration (f) good energy resolution,
typically of the order of 10\% or better for the heavier nuclei at
$\gamma \approx$ 1000.

Traditionally, a TRD uses multiwire proportional chambers (MWPC's) as
x-ray detectors. MWPC's require the use of a pressurized container, as
employed in the Cosmic Ray Nuclei (CRN) experiment which was the first
TRD flown in space [\refcite{lheureux}]. The weight of such a
container is large, and the need for it can be avoided by the approach
taken by TRACER. For the TRACER instrument the MWPC's are replaced
with layers of thin-walled single wire proportional tubes which can
easily withstand internal overpressure.

\subsection{The TRACER Instrument}

Balloon borne experiments are limited in size and weight. Payloads are
presently limited to $\sim$2700~Kg, of which $\sim$700~Kg is reserved
for satellite telemetry interfaces, ballast and landing gear
(parachute and crush pad). TRACER is at the upper limit in terms of
size and weight for a balloon payload. The employment of a TRD,
without the need of a pressurized container allows for the
construction of a large area detector (2.06~x~2.06~m$^{2}$), resulting
in an overall geometric factor of 5~m$^{2}$sr. The individual
components of TRACER are shown in Figure~\ref{tracer} and are
summarized here. From top to bottom the instrument consists of:

\begin{itemize}

\item[1] \emph{Scintillator Counter}: A plastic counter (BICRON-408)
that has an active area of 2~m x 2~m and is 0.5 cm thick. The counter is
read out via wavelength shifter bars (BICRON 482) by 24
photomultiplier tubes (Photonis XP1910).

\item[2] \emph{dE/dx Counter}: An array of 800 single wire proportional
tubes. Each tube is 2~m long and 2cm in diameter and the walls, which
are made of mylar, are 127 $\mu$m thick. The tubes are arranged in a
total of four double layers, with each layer consisting of 100
tubes. Each double layer is orientated in an orthogonal direction to
the one above.

\item[3] \emph{Transition Radiation Detector}: A second array of 800
single wire proportional tubes identical to the dE/dx counter except
that each double layer is preceded by blankets of plastic fiber
material which act as a radiator to generate transition radiation.

\item[4] \emph{Scintillator Counter}: A second counter of identical
design as the top scintillator counter is placed below the TRD.

\item[5] \emph{Cherenkov Counter}: An acrylic counter with refractive
index = 1.49 is placed at the bottom of the stack. The active area of
the counter is 2~m x 2~m and the thickness is 1.27~cm. The counter is
also read out, via wavelength shifting bars, by 24 PMTs.

\end{itemize}

\begin{figure}[ht]
\begin{center}
\includegraphics[width=0.95\textwidth]{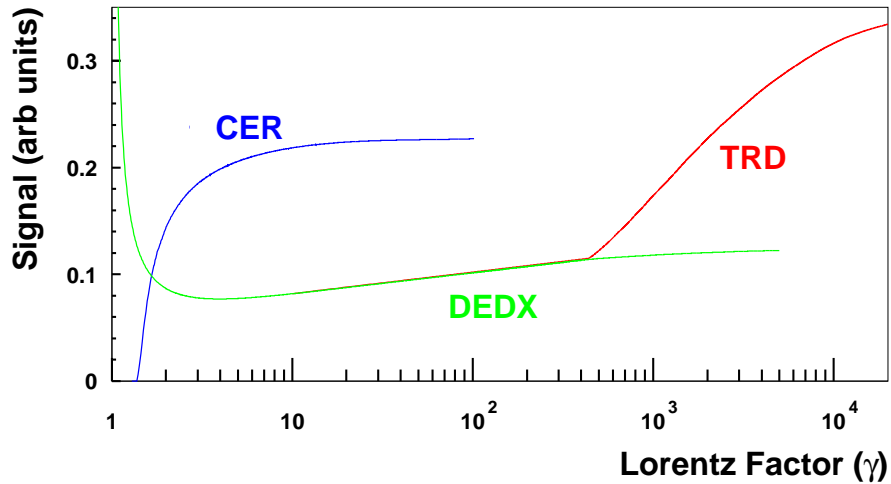}
\end{center}
\caption{Energy scale of TRACER. TRACER combines three energy
measurements to cover more than 4 decades in energy. The relative
amplitude of the response curves is representative of the
signal/$Z^{2}$ in each detector [From Ave et al.]}
\label{esig}
\end{figure}

The scintillator counters act as a trigger for the instrument and
together with the Cherenkov detector measure the nuclear charge
\emph{Z} of each individual cosmic-ray particle traversing the
instrument. The array of all 1600 proportional tubes determines the
trajectory of each cosmic-ray particle through the instrument. The
energy response of the TRACER detector spans 4 decades in energy and
is shown in Figure~\ref{esig}. This extensive energy range is
accomplished by combining three complementary measurements: Cherenkov
light measurement ($\sim 10^{11}$~eV), the relativistic rise of the
ionization signal in gas ($\sim 10^{11}-10^{13}$~eV) and the
measurement of transition radiation ($> 10^{13}$~eV).

\subsection{Balloon Flights}

TRACER has had three successful flights on high altitude balloon thus
far. The instrument and data have been recovered intact after each
flight. A summary of the flights is as follow:

\emph{\underline{Flight I}}: Launch from Ft. Sumner, USA on September
20, 1999 on a 39 million cubic-foot balloon. The flight was 28 hours
in duration at a float altitude between 34 and 38 km, corresponding to
a residual atmospheric depth of 4-6.5 $\mathrm{g\:cm^{-2}}$. This
flight served as a test flight for the subsequent longer duration
flights. 

\emph{\underline{Flight II}}: Launch from Antarctica on December 12,
2003 on a 39 million cubic-foot balloon. The flight lasted 14 days at
a float altitude between 36 and 39 km, corresponding to an average
residual atmospheric depth of 3.9 $\mathrm{g\:cm^{-2}}$. A total of
5~$\times$~10$^7$ cosmic-ray particles were collected. 

For the first two flights, the trigger threshold was set such that the
instrument had full efficiency for the elements oxygen to iron.

\emph{\underline{Flight III}}: Launch from Kiruna, Sweden on July 8,
2006 on a 39 million cubic-foot balloon. The flight was limited to 4.5
days due to the lack of permission to fly over Russian territory. The
float altitude was between 36 and 40 km, corresponding to an average
residual atmospheric depth of 3.5 $\mathrm{g\:cm^{-2}}$. For this
flight, the trigger threshold was set at a signal level corresponding
to a nuclear charge between lithium and beryllium. This allowed for
full efficiency for the lighter elements boron, carbon, and nitrogen
in addition to the heavier elements. A total of 3~$\times$~10$^7$
cosmic-ray particles were collected.

The results of \emph{Flight I} have been published by Gahbauer et
al. (2004) [\refcite{gahbauer03}]. The subsequent discussion of the
analysis and results from TRACER presented in this paper pertain to
\emph{Flight II}. These results have been published by Ave et
al. [\refcite{ave08}], where the analysis is discussed in greater
detail. Analysis of data collected in \emph{Flight III} is currently
ongoing.

\section{Constructing an Energy Spectrum with TRACER}
\label{sec:analysis}
The analysis of the TRACER data begins with the reconstruction of the
trajectory of each cosmic-ray particle through the instrument. The
accurate knowledge of the particle trajectory is essential to the
analysis for two reasons; first, it permits corrections of the
scintillator and Cherenkov signals due to spatial non-uniformities and
zenith-angle variations, second, it makes it possible to determine
accurately the energy per unit pathlength deposited by each cosmic ray
traversing the proportional tubes.

\subsection{Trajectory Reconstruction}
The trajectory of cosmic rays through the instrument is reconstructed
for signals recorded in the entire proportional tube array. The
reconstruction follows a two step procedure. As a first estimate, the
trajectory is obtained by constructing a straight line fit to the
center of each of the tubes hit in an event. All possible combinations
of these tubes are fit in the x- and y-projections, and the
combination with the minimum $\chi^{2}$ per degree of freedom and
maximum number of tubes hit is kept. This procedure reconstructs the
trajectories of over 95\% of all cosmic-ray particles, with a lateral
accuracy in the track position of 5~mm. The trajectory is then refined
using the fact that the energy deposited in each tube is proportional
to the track length in that tube. Taking this fact into account, an
accuracy of 2~mm in lateral track position is achieved, which
corresponds to a 3\% uncertainty in the total path length through all
tubes.

\subsection{Charge Reconstruction}
\begin{figure}[ht]
\begin{center}
\includegraphics[width=0.49\textwidth]{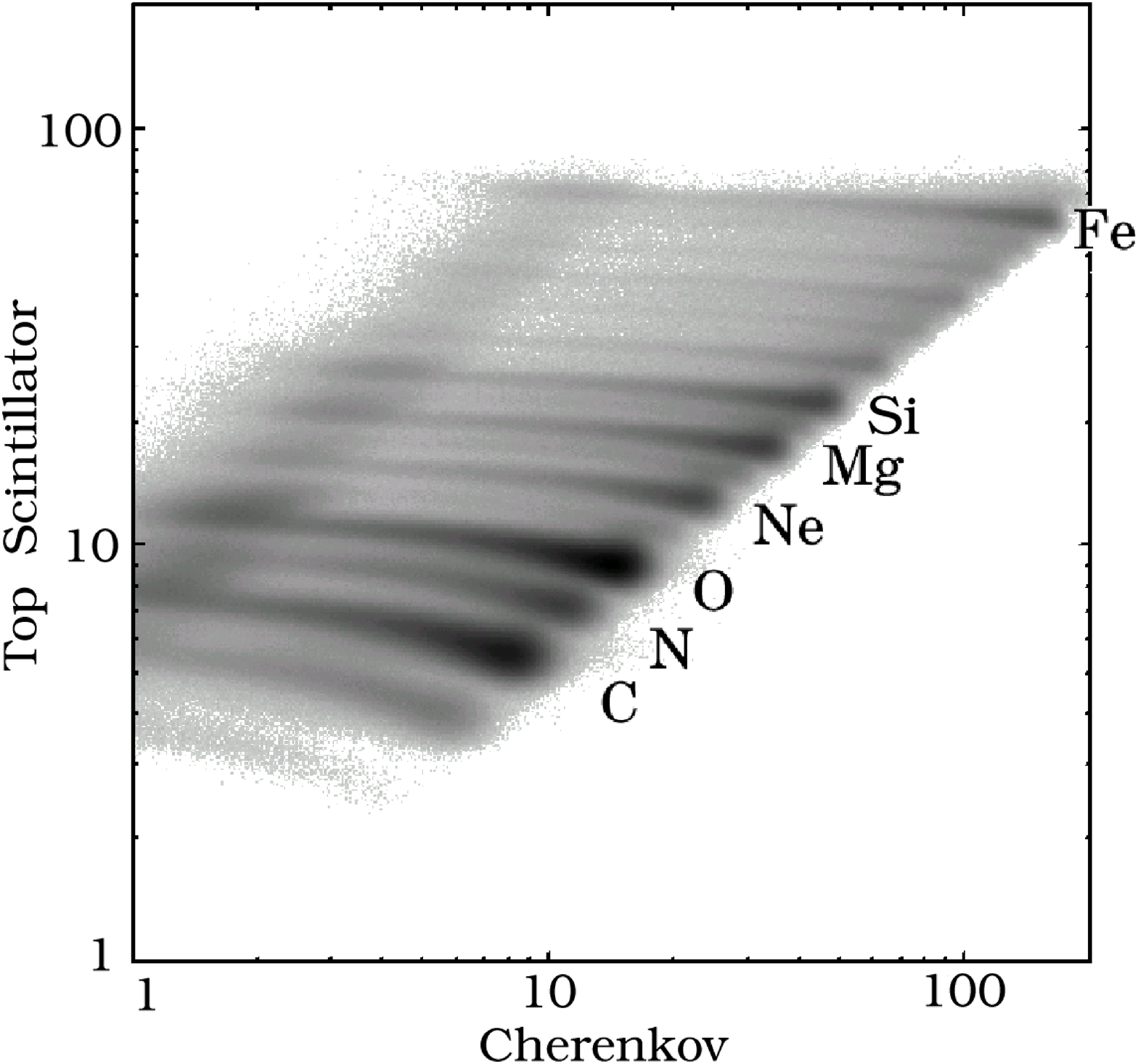}
\includegraphics[width=0.47\textwidth]{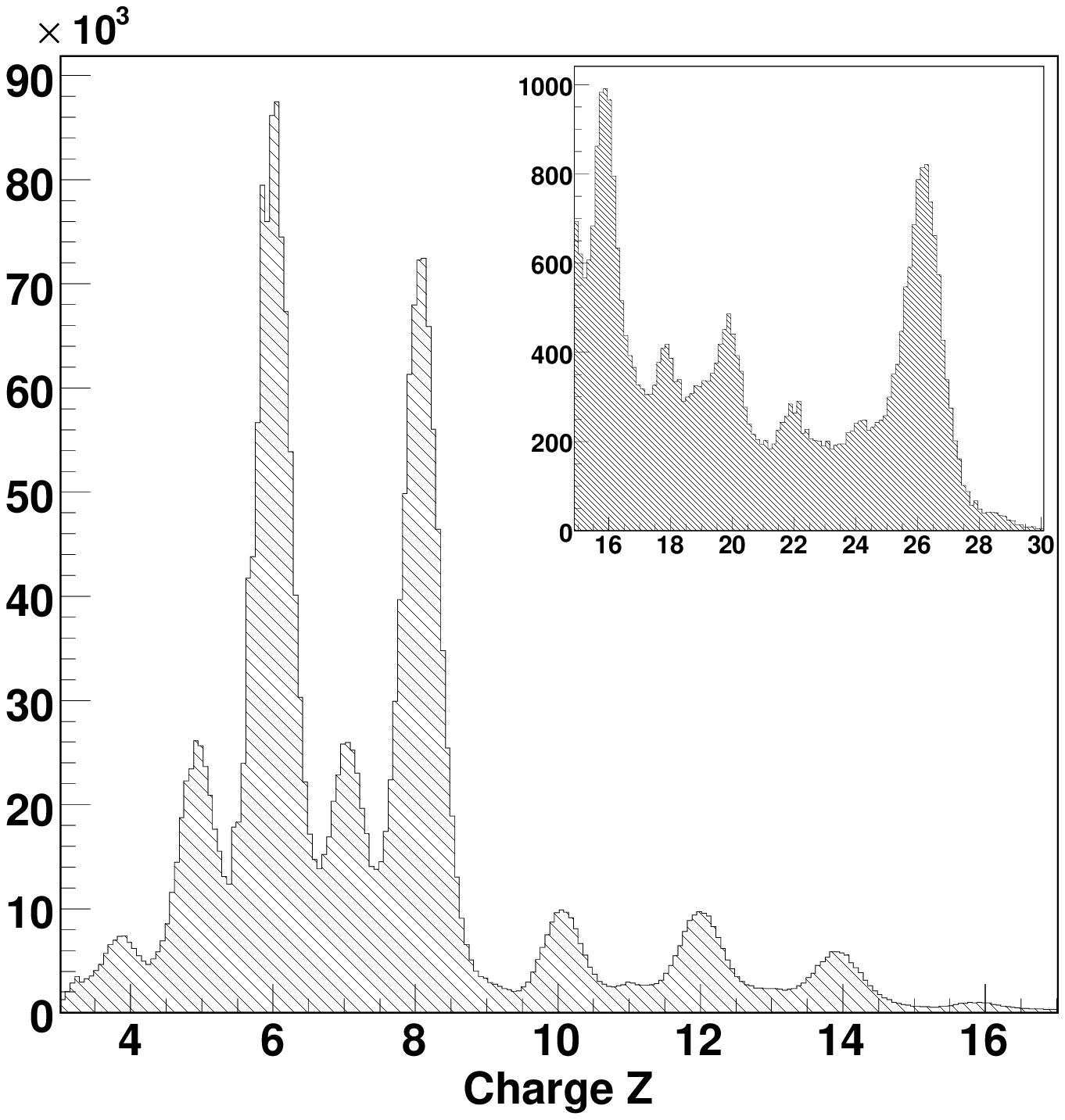}
\end{center}
\caption{Scatter plot of the signal from the scintillator and
Cherenkov counters (left). A charge histogram for all events is
obtained by summing along lines of constant charge (right). [From Ave
et al.]}
\label{charge}
\end{figure}

The charge is determined for each cosmic-ray particle using the
signals from the scintillation and Cherenkov counters. The signals are
corrected for spatial non-uniformities in the counter responses using
the trajectory information together with response maps recorded with
muons on the ground. Figure~\ref{charge} shows a correlation of the
signals from the top scintillator and Cherenkov counters. Particles of
constant charge are clustered along the lines, with the position on
the line dependent on the primary energy. A charge histogram is then
constructed through the summation along these lines of constant
charge.

The resulting charge resolution is both charge and energy
dependent. For oxygen nuclei (Z=8) the charge evolves from 0.25 at
energies below 3 GeV amu$^{-1}$ to 0.30 charge units at higher
energies, while for iron nuclei (Z=26) the resolution is 0.5 and 0.6
charge units respectively.

\subsection{Energy Reconstruction}
The energy of each cosmic-ray particle is obtained from the combined
signals of the Cherenkov counter and of the proportional tubes. The
Cherenkov counter identifies particles below minimum ionization energy
(3~GeV~amu$^{-1}$), and also provides a good energy measurement for
these low-energy particles (see Figure~\ref{esig}). Between minimum
ionization energy and the onset of TR (400~GeV~amu$^{-1}$), the
signals in the dE/dx counter and the TRD are the same and increase
logarithmically with energy (see Figure~\ref{esig}).  Above 400 GeV
amu$^{-1}$, the signals from the dE/dx counter and TRD diverge, and
the rapid increase in the TR signals with particle energy provides an
excellent energy measurement. However, cosmic-ray particles in this
energy region are extremely rare, being less abundant than particles
in the minimum ionization energy region by more than 4 orders of
magnitude. To uniquely identify these rare cosmic-ray particles, it is
required that particles with low-energy are identified and removed on
the basis of their Cherenkov signals. To select the highest energy
events, it is required that the measurement in the dE/dx counter
places these particles at an energy level well above minimum
ionization energy and the presence of TR is detected.  Thus the
combination of the Cherenkov counter, the dE/dx counter and the TRD is
crucial for the success of the TRACER measurement at the highest
energies.

\begin{figure}[ht]
\begin{center}
\includegraphics[width=0.65\textwidth]{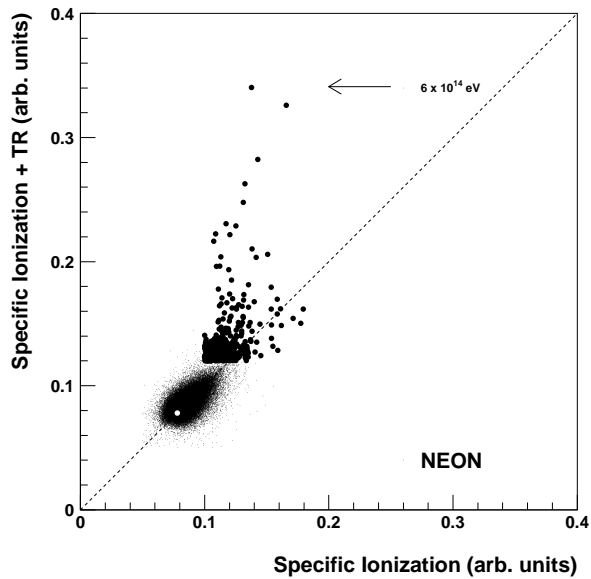}
\end{center}
\caption{Scatter plot of the signal from the TRD and the DEDX counters
for neon nuclei. The highlighted points represent the highest energy
events as measured with the TRD. As expected the TR events have
signals in the DEDX counter which are well above the minimum
ionization energy (white circle). [From Ave et al.]}
\label{neon}
\end{figure}

Figure~\ref{neon} shows a cross-correlation plot between the TRD and
dE/dx counter signals for neon nuclei (Z=10). Cosmic-ray particles
below minimum ionization energy have been removed with the aid of the
Cherenkov counter. The small black points represent particles with
energies below the onset of TR. As the signals from the TRD and dE/dx
counter are similar in this energy regime, the correlation of the
signals lies along the diagonal. At the onset of TR, where the TRD and
the dE/dx signals diverge, the highest energy events show up away from
the diagonal line. The off-diagonal position of these highest energy
particles defines them uniquely as TR events for this selected
charge. The highest energy neon nucleus in this data sample has an
energy of $6 \times 10^{14}$~eV~particle$^{-1}$.

\subsection{Absolute Intensities}

Once each cosmic-ray particle has been assigned an energy it is sorted
into energy bins of width $\Delta$E$_{i}$ and a differential energy
spectrum is constructed for each elemental species. To convert from
the number of events $\Delta$N$_{i}$ in a particular energy bin
$\Delta$E$_{i}$ to an absolute flux \emph{dN/dE(i)} one must compute
the exposure factor, effective aperture of the instrument, and the
efficiency of the analysis selection and instrument response. The
overall efficiency is high, with a tracking efficiency of $\sim$95\%,
livetime $\sim$94\%, and charge selection efficiency 70-80\% (for the
more abundant elements). Further details of the reconstruction
efficiencies can be found in Ave et al. [\refcite{ave08}].

\begin{figure}[ht]
\begin{center}
\includegraphics[width=0.95\textwidth]{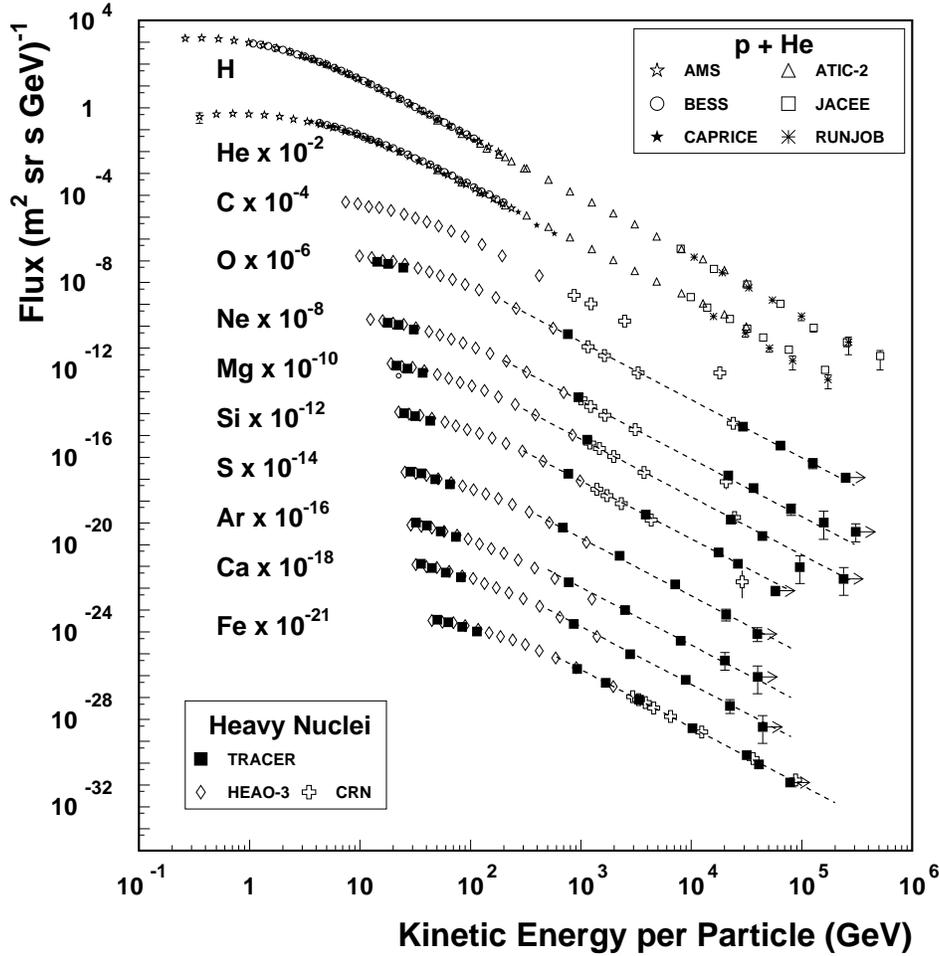}
\end{center}
\caption{Flux as a function of energy for the major components of the
primary cosmic radiation. The measurements by the TRACER experiment
are represented by the solid squares for the elements O, Ne, Mg, Si,
S, Ar, Ca, and, Fe. The dashed line represents a power-law fit to the
TRACER data above 20~GeV~amu$^{-1}$. For references to the data
presented in this plot see [7]-[16] and references therein.}
\label{spectra}
\end{figure}

\section{Resulting Energy Spectra}
\label{sec:results}
The energy spectra, in terms of absolute intensities at the top of the
atmosphere, for the elements O, Ne, Mg, Si, S, Ar, Ca, and Fe are
plotted as solid squares in Figure~\ref{spectra}. For clarity, the
intensity of each element is scaled by a factor shown on the
left. Existing data from measurements in space with HEAO-3 (open
diamonds [\refcite{heao}]) and CRN (open crosses [\refcite{crn}]) are
shown for comparison. For completeness, data on the light primary
cosmic ray components (protons, helium, and carbon) which are not
measured with TRACER are also included. These data come from
measurements in space (AMS [\refcite{ams}]) and on balloons (ATIC
[\refcite{atic}], BESS [\refcite{bess1}, \refcite{bess2}], CAPRICE
[\refcite{caprice}], JACEE [\refcite{jacee}], RUNJOB
[\refcite{runjob}]). 

Note the large range in intensity (10 decades) and particle energy (4
decades) covered by TRACER. As can be seen, the energy spectra for O,
Ne, Mg, and Fe extend up to and beyond 10$^{14}$ eV
particle$^{-1}$. The energy range is limited by the current exposure
and not by the saturation of the TRD. No significant change in
spectral slope is evident at the highest energies. The energy spectrum
of each element, as measured by TRACER, can be fit to a power law
above 20 GeV amu$^{-1}$ (dashed line in Figure~\ref{spectra}). The
resulting spectral indices are remarkably similar with an average
exponent of -2.65 $\pm$ 0.05.

\begin{figure}[ht]
\begin{center}
\includegraphics[width=0.65\textwidth]{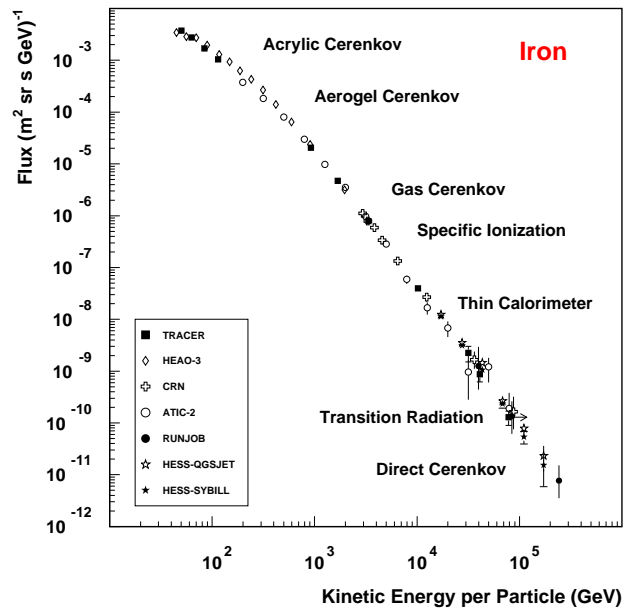}
\end{center}
\caption{Energy spectrum for iron nuclei highlighting the
complementarity between detection techniques employed in cosmic-ray
composition measurements.}
\label{iron}
\end{figure}

Figure~\ref{iron} compares the TRACER results for iron nuclei with
results from previous experiments in space (HEAO-3~[\refcite{heao}]
and CRN~[\refcite{crn}]), and on balloons (ATIC-2~[\refcite{atic}] and
RUNJOB~[\refcite{runjob}]). Figure~\ref{iron} also illustrates the
variety of detection techniques used in measuring the energy of heavy
nuclei. Within the statistical uncertainties (which in some
measurements are quite large), the data indicate fairly consistent
results.  Also presented in Figure~\ref{iron} are recent results from
the ground based HESS Imaging Air Cherenkov Telescope using the Direct
Cherenkov Technique [\refcite{kieda}]. Here, two flux values are
presented for each energy arising from ambiguities from different
nuclear interaction models used in the data analysis
[\refcite{hess}]. Again, the HESS results are consistent with TRACER.

\begin{figure}[ht]
\begin{center}
\includegraphics[width=0.95\textwidth]{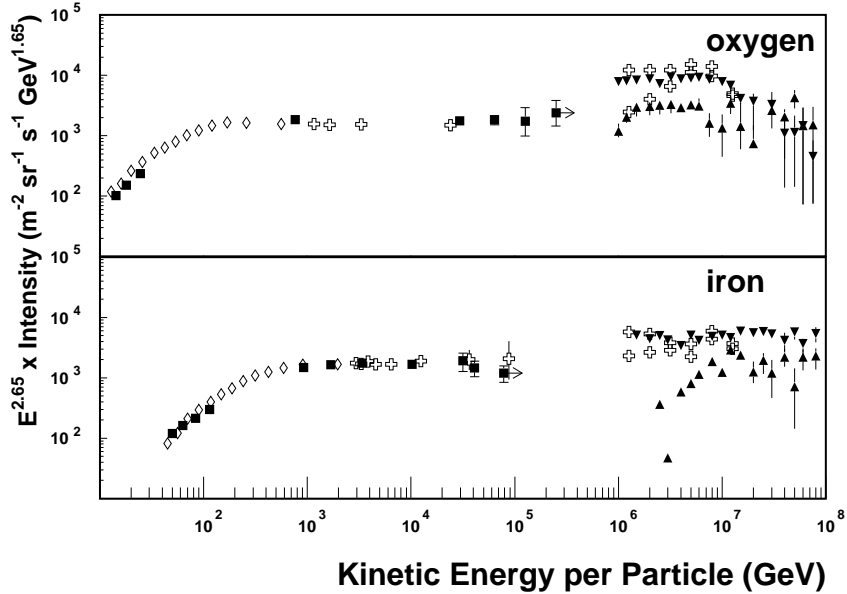}
\end{center}
\caption{Energy spectra multiplied by E$^{2.65}$ from TRACER (solid
squares) compared with the interpretation of air-shower data of
KASCADE (filled triangles, for two different interaction models:
Antoni et al. (2005)) and of EAS-TOP (open crosses, two data points
for each energy represent upper and lower limits: Navarra et
al. (2003).  [From Ave et al.]}
\label{airshower}
\end{figure}

\subsection{Comparison with Air-shower Data}

Data for oxygen and iron nuclei from TRACER are compared in
Figure~\ref{airshower} with energy spectra derived from indirect
air-shower observations of the EAS-TOP collaboration
[\refcite{eastop}] and of the KASCADE group [\refcite{kascade}]. These
experiments report results for groups of elements and not individual
elements; therefore the fluxes for the ``CNO group'' probably have
about twice the intensity than oxygen alone while the ``iron group''
is probably dominated by iron. Furthermore, these results remain
ambiguous as they depend strongly on the choice of the nuclear
interaction model used in the analysis. The TRACER results do not yet
overlap with the energy region of the air-shower data. Additional
measurements with TRACER should help to close the gap and to provide
significant constraints on the interpretation of air-shower results.

\section{Future Measurements with TRACER}
\label{sec:future}

Extending the energy spectra of the heavy elements towards
10$^{15}$~eV~particle$^{-1}$ is an ongoing goal of TRACER. However, to
determine the energy spectra and relative abundances for each nuclear
species at the cosmic-ray source, the mode of galactic propagation
must be understood. The simplest model of cosmic-ray propagation
assumes an equilibrium between the production of cosmic rays, from
acceleration at the source or from spallation of heavier nuclei, and
their loss from competing actions of escape from the Galaxy and
spallation on the interstellar medium [\refcite{swordy93}]. This model
can be described as:

\vspace{-0.1cm}
\begin{center}
\begin{equation}
\label{leakybox}
N_{i}(E) = \frac{1}{\Lambda(E)^{-1} + \Lambda_{s}^{-1}}(\frac{Q_{i}(E)}{\beta c \rho} + \sum_{k > i} \frac{N_{k}}{\Lambda_{k \to i}})
\end{equation}
\end{center}

\noindent
where $N_{i}(E)$ is the observed intensity of element $i$,
$\Lambda(E)$ the propagation pathlength, $\Lambda_{s}$ the spallation
pathlength, $Q_{i}(E)$ the rate of production in the cosmic-ray
source, $\beta = v/c$, $\rho$ the average density of the interstellar
medium, $N_{k}$ the intensity of element $k$, and $\Lambda_{k \to i}$
the spallation mean free path for an element $k$ to spallate to
element $i$.


\begin{figure}[ht]
\begin{center}
\includegraphics[width=0.65\textwidth]{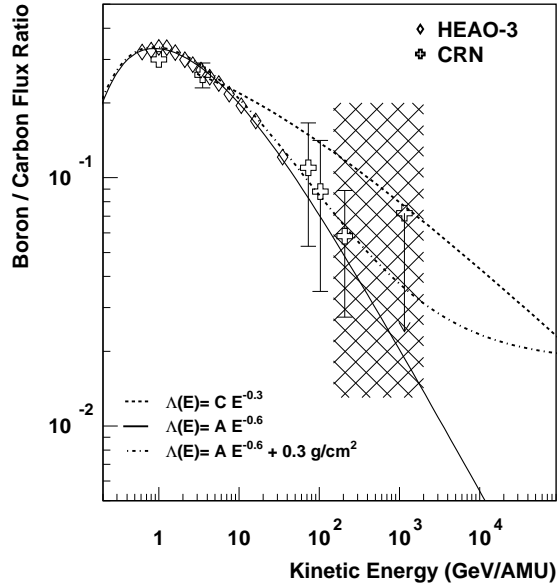}
\end{center}
\caption{Boron to carbon flux ratio with data from HEAO-3 and CRN. The
solid line represent a parameterization given by Yanasak. The addition
of a residual pathlength of $\Lambda_0$ = 0.3 g/cm$^{2}$ to the
Yanasak parameterization is presented by the dash-dot line. The dashed
line represents a fit to $\Lambda(E)~\propto~E^{-0.3}$. The hashed
area represents the energy region attainable from the 5-day flight of
TRACER in 2006.}
\label{bc}
\end{figure}

The energy dependent propagation pathlength $\Lambda(E)$ can be
derived from the flux ratio of secondary spallation produced particles
to primary accelerated particles. This flux ratio has been observed to
decrease with increasing energy [\refcite{juliusson}], indicating that
the mean lifetime or propagation pathlength of cosmic rays in the
Galaxy decreases with energy. As an example, the boron to carbon flux
ratio as measured by HEAO-3 [\refcite{heao}] and CRN [\refcite{crn}]
is shown in Figure~\ref{bc}. The flux ratio has been parameterized
with $\Lambda(E) \propto E^{-0.6}$ [\refcite{heao}]. However, this
flux ratio has been measured accurately only to about 10$^{12}$~eV
particle$^{-1}$. It may well be that the flux ratio will eventually
reach a finite asymptotic value $\Lambda_0$
(i.e. $\Lambda(E)~=~AE^{-0.6}~+~\Lambda_0$). The residual pathlength
$\Lambda_0$ would represent the minimum amount of matter high-energy
particles must encounter during propagation from the cosmic-ray source
to Earth. Alternatively, some diffusion propagation models expect a
Kolmogorov spectrum with $\Lambda(E) = CE^{-0.3}$ (see
[\refcite{strong}] for a comprehensive review, and references
therein). Figure~\ref{bc} illustrates that none of these scenarios
cannot be fully ruled out with the present data. An extension of the
boron-carbon flux ratio was a major science goal of \emph{Flight III}
of TRACER in 2006.  It is expected that even with the limited exposure
obtained in \emph{Flight III}, the measurement of the boron-carbon
flux ratio will reach into the 10$^{13}$-10$^{14}$~eV~particle$^{-1}$
range. This energy region is represented as the shaded area in
Figure~\ref{bc}.

\begin{figure}[ht]
\begin{center}
\includegraphics[width=0.75\textwidth]{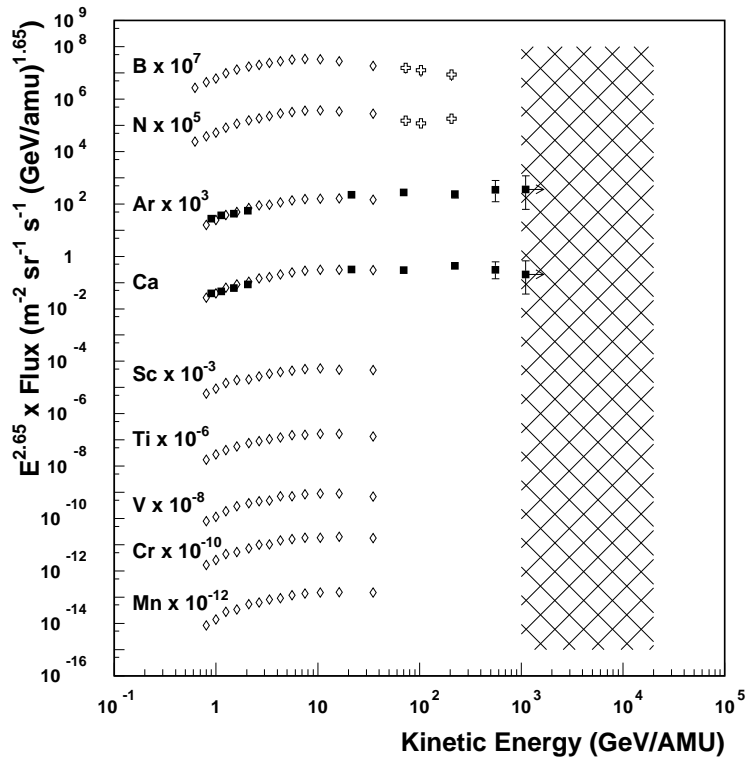}
\end{center}
\caption{Energy spectra multiplied by E$^{2.65}$ of purely secondary
elements (B and sub-Fe) and mixed elements (primary and secondary: N,
Ar, and Ca) from the HEAO-3 (open diamonds), CRN (open crosses) and
TRACER (solid squares). The hashed region represents the energy region
attainable with a 30 day flight of the upgraded TRACER.}
\label{future}
\end{figure}

The flux ratio of the secondary sub-Fe elements ($Z$=21-25) to iron
($Z$=26) can also be used to determine the propagation parameters
$\Lambda(E)$ and $\Lambda_{0}$. The sub-Fe elements are produced by
the spallation of iron nuclei during propagation. Figure~\ref{future}
shows the currently available energy spectra for the sub-Fe elements
along with B, C, Ar, and Ca spectra. The energy spectra for the sub-Fe
elements do not yet reach 10$^{12}$~eV~particle$^{-1}$.

\begin{figure}[ht]
\begin{center}
\includegraphics[width=0.49\textwidth]{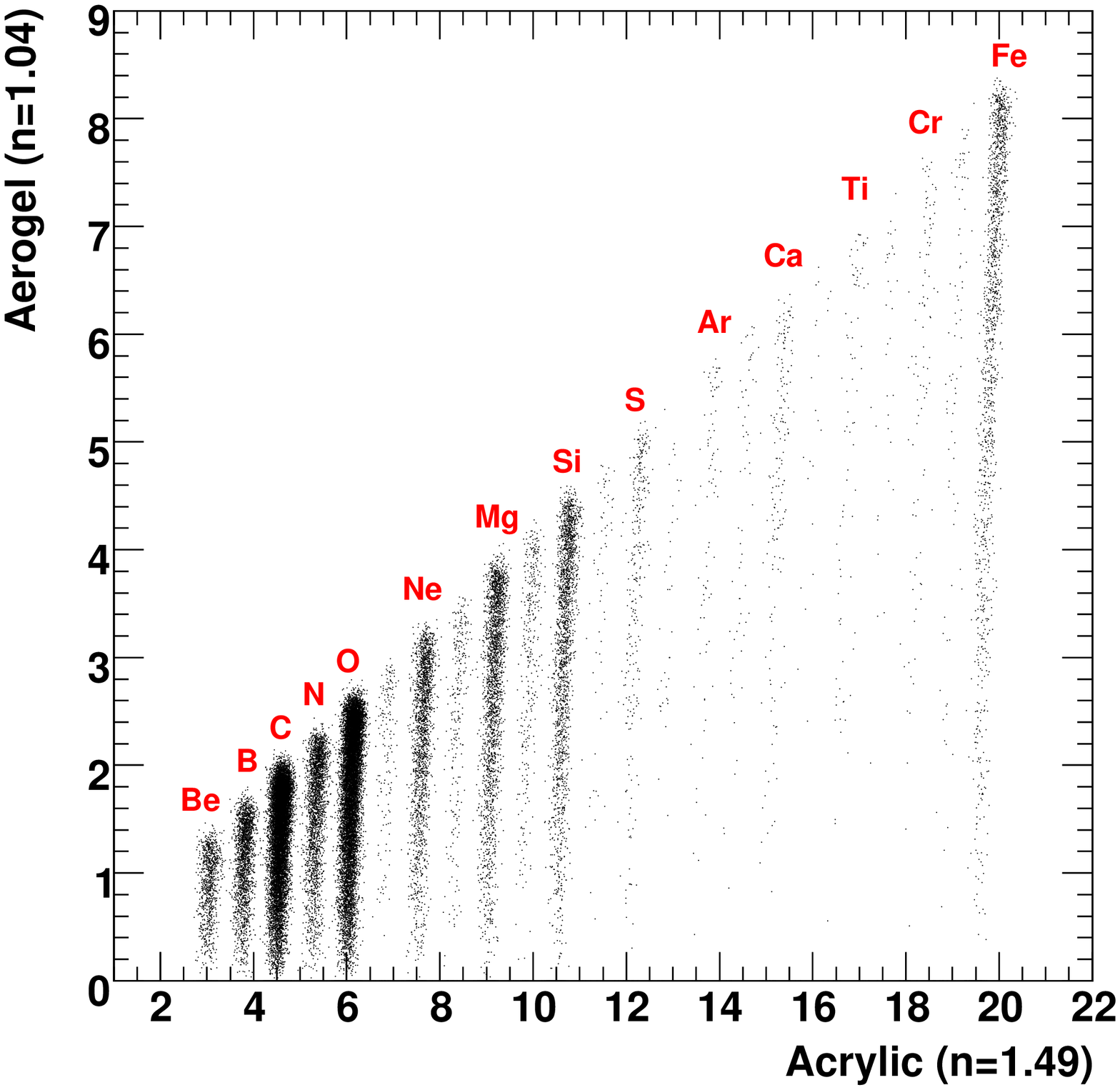}
\includegraphics[width=0.48\textwidth]{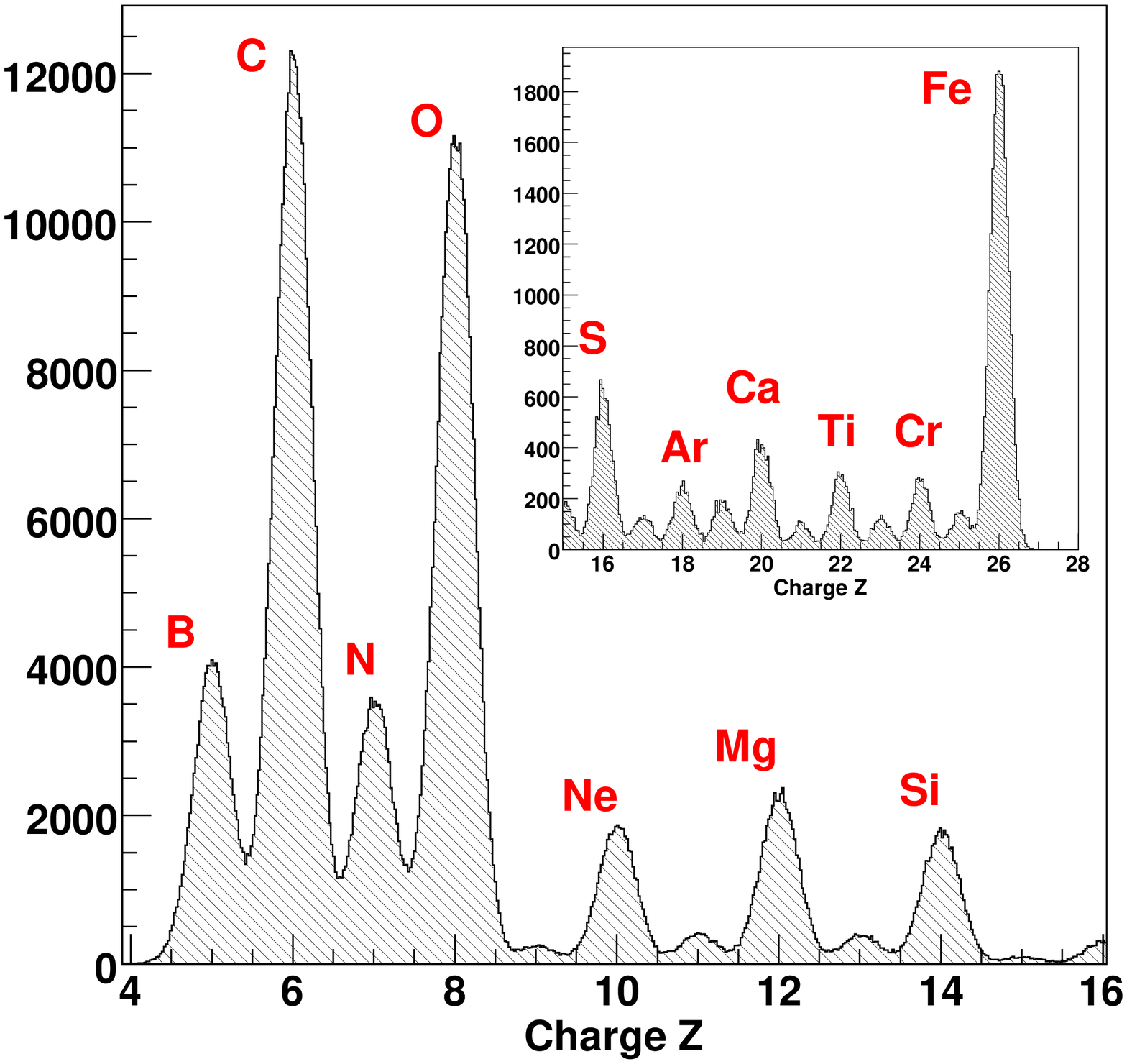}
\end{center}
\caption{(Left-hand panel) Cross correlation plot of simulated signals from
two Cherenkov counters; one containing an aerogel radiator (refractive
index = 1.04) and the second an acrylic plastic radiator (refractive
index = 1.49). (Right-hand Panel) Resulting charge histogram with charge
resolution of 0.2 charge units.}
\label{newcharge}
\end{figure}

TRACER is unable to measure the energy spectra for the sub-Fe elements
with the present detector configuration. The main reasons for this are
the limited charge resolution, due to nonlinearity of the scintillator
response at high charges, and the limited photo-electron statistics of
the Cherenkov counter. It is therefore proposed to replace one of the
acrylic Cherenkov counters with a counter containing an aerogel
radiator (refractive index 1.04), and to use a more efficient readout
system for the second acrylic Cherenkov counter. For relativistic
particles the charge would then be reconstructed using the acrylic and
the aerogel Cherenkov counters. Contrary to the behavior of the
scintillators, both signals are strictly proportional to $Z^{2}$ and
the charge resolution would be independent of charge. This technique
has been successfully utilized in the balloon borne Trans Iron
Galactic Recorder (TIGER) instrument [\refcite{link}], which achieved
a charge independent resolution of 0.22 charge units. Monte-Carlo
simulations have shown that a charge resolution of 0.2 charge units
could be achieved with the proposed Cherenkov counter system
[\refcite{marshall},\refcite{hams}]. Figure~\ref{newcharge} shows the
results from a simulated charge reconstruction for such a setup. The
left-hand panel shows a cross correlation plot of the simulated
signals from the aerogel and acrylic counters. The vertical lines
represent cosmic rays of constant charge. The resulting charge
histogram, shown in the right-hand panel of Figure~\ref{newcharge},
indicates that the sub-Fe elements are clearly resolved. As Antarctic
flights of 30 days duration are now becoming a reality
[\refcite{jones}], such an exposure of the upgraded TRACER instrument
would extend the energy spectra for all secondary elements to
10$^{14}$~eV~particle$^{-1}$. This energy region is indicated by the
shaded region in Figure~\ref{future}.

\section{Conclusions}

The TRACER program has pioneered the development of a TRD using
proportional tube arrays to directly measure the energy spectra of
cosmic-ray nuclei. The coupling of the TRD with the Cherenkov and DEDX
counters enables an energy measurement of individual nuclei across 5
decades in energy and provides a technique to uniquely identify the
rare high-energy particles with a discrimination power of $> 10^{4}$.

TRACER has measured the energy spectra of the elements O, Ne, Mg, Si,
S, Ar, Ca, and Fe up to 10$^{14}$~eV~particle$^{-1}$. This data set
currently represents the most comprehensive measurements to date of
heavy nuclei with individual charge resolution. The measured energy
spectra reach similar power laws at high energies and no change in
slope is evident. The data indicate a common origin and mode of
propagation for all elements and support the SNR theory of the origin
of galactic cosmic rays.

TRACER has been upgraded to include a measurement of the secondary to
primary flux ratio. Data analysis of the boron to carbon flux ratio is
currently underway. Future upgrades of TRACER will include
measurements of the sub-Fe elements.

\section*{Acknowledgments}

This work has been made possible by the enormous contributions of the
TRACER team: Maximo Ave, Florian Gahbauer, Christian H\"{o}ppner,
J\"{o}rg H\"{o}randel, Masakatsu Ichimura, Jesse Marshall, Dietrich
M\"{u}ller, Andreas Obermeier, and Andrew Romero-Wolf. We gratefully
acknowledge the services of the University of Chicago Engineering
Center in particular Gary Kelderhouse, Gene Drag, Casey Smith, Richard
Northrop and Paul Waltz. We thank the staff of the Columbia Scientific
Balloon Facility, the NSF Antarctic Program and the Swedish Space
Corporation for support during the balloon campaigns. This work was
supported by NASA grants NAG55305, NN04WC08G and NNG06WC05G. Numerous
students have participated in the construction and refurbishment of
the instrument under support from the Illinois Space Grant Consortium.


\end{document}